\def\imat {\rm i}

\documentstyle[twocolumn,floats,epsf,ifthen,aps,prb]{revtex} 

\begin{document} 

 

\title{Quantum feedback control of a solid-state two-level system} 

\author{Rusko Ruskov\footnote{On leave of absence from 
Institute of Nuclear Research and Nuclear Energy, Sofia BG-1784, Bulgaria} 
and Alexander N. Korotkov\footnote{Electronic mail: korotkov@ee.ucr.edu}} 
\address{ 
Department of Electrical Engineering, University of California, 
Riverside, CA 92521-0204. 
} 
\date{\today} 
 
\maketitle 
 
\begin{abstract} 
We have studied theoretically the basic operation of a quantum feedback loop 
designed to maintain the desired phase of quantum coherent oscillations 
in a two-level system. Such feedback can suppress 
the dephasing of oscillations due to interaction with environment. 
Prospective experiments can be realized using metallic single-electron 
devices or GaAs technology. 
\end{abstract} 
 
\narrowtext 
 
\vspace{0.6cm} 

        The principle of feedback control is used in a wide variety
of physical and engineering problems. In particular, it can be applied
in a straightforward way to tune the oscillation phase of a harmonic 
oscillator in order to achieve a desired synchronization with some 
reference oscillator. An intriguing and fundamental question is whether
continuous feedback can be used to control quantum systems; for instance, 
if it is possible or not to tune the phase of quantum coherent (Rabi)
oscillations in a two-level system (TLS). 

        At first sight the quantum feedback seems to be impossible
because according to the ``orthodox'' collapse postulate \cite{Neumann}  
the quantum state is abruptly destroyed by the act of measurement. 
However, as was shown two decades ago, in particular by Leggett,  
\cite{Caldeira} in a typical solid-state setup the collapse of a 
TLS state should be considered as a continuous process rather
than as instantaneous event. The reason is typically weak coupling
between the quantum system and the detector and also the finite noise
of the detector, so that it takes some time until acceptable 
signal-to-noise ratio is reached and the measurement can be regarded  
as completed. 

        While the Leggett's theory as well as the majority of similar 
approaches can describe only {\it ensembles} of quantum systems, 
the theory describing the gradual collapse of a {\it single} solid-state 
TLS was developed only recently.\cite{Kor-PRB,Kor-meas2,Goan} 
(A similar problem in optics was solved much earlier -- see, e.g. 
Refs.\ \cite{Carmichael,Wiseman} and references in \cite{Kor-meas2}.) 
Basically, the theory says that the evolution of a single quantum system 
due to continuous measurement is governed by the information continuously 
acquired from the detector. Similarly to classical probability,
the Bayes formula \cite{Feller} which naturally takes into account incomplete 
information from the detector, can still be applied to the density matrix
of the measured quantum system; thus the formalism is called 
Bayesian.\cite{Kor-PRB} 

        In case of a poor detector the extra noise acting on the input 
disturbs the measured system stronger than the limit determined by 
the uncertainty principle; this leads to gradual decoherence of the 
measured system. In contrast, when measured with a good (quantum-limited) 
detector, the quantum system does not loose the coherence (even though
the quantum state evolves randomly); moreover, 
its density matrix can be gradually purified\cite{Kor-PRB} 
that basically means acquiring as much information about the system 
as permitted by quantum mechanics. 

        Since the Bayesian formalism allows us to monitor the continuous
evolution of a quantum system in a process of measurement, this naturally 
gives rise to a possibility of continuous feedback control of a quantum
system. In this paper we will study the operation of a feedback loop 
proposed in Ref. \cite{Kor-meas2} and designed to maintain a desired 
phase of quantum coherent oscillations in a solid-state TLS. 
(Quantum feedback in optics has been proposed and studied earlier 
-- see, e.g., Refs. \cite{Wiseman,Tombesi,Doherty}.)  
In particular, we will study the dependence of the loop operation
on the feedback factor $F$, the available bandwidth $\tau_a^{-1}$, 
and the dephasing rate $\gamma_e$ due to environment. 

\begin{figure} [t] 
\centerline{
\epsfxsize=2.7in 
\epsfbox{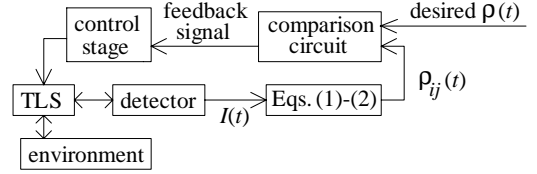}
} 
\vspace{0.4cm} 
\caption{ Schematic of the quantum feedback loop maintaining the 
quantum oscillations in a two-level system (TLS). 
 }
\label{schematic}\end{figure}

        As an example of the measurement setup (Fig.\ \ref{schematic}) 
we consider a TLS represented by a single electron in a  
double quantum dot (DQD), the location of which is measured by 
a quantum point contact (QPC) nearby in a way used in Ref.\ \cite{Buks}. 
If the electron is in the dot 2 (state $|2\rangle$) which is closer  
to QPC than dot 1, then the QPC tunnel barrier is higher and so 
the average current $I_2$ through QPC is smaller than the average current 
$I_1$ corresponding to electron in the dot 1 (state $|1\rangle$). 
Consequently, from the QPC current one gets information about the 
electron location. 
We consider a realistic case of weak response, 
$\Delta I\equiv I_1-I_2 \ll I_0\equiv (I_1+I_2)/2$.  
In this case the measurement time $S_I/2(\Delta I)^2$, which is necessary
to achieve signal-to-noise ratio equal to 1 (here $S_I$ is the shot 
noise of the QPC current), is much larger than $e/I_0$, so the QPC current 
$I(t)$ is continuous on the measurement timescale and we do not need 
to consider individual tunneling events in QPC. 

        The evolution of the TLS density matrix $\rho$ during the 
measurement process is described within the Bayesian formalism 
by equations \cite{Kor-PRB,Kor-meas2}  
        \begin{eqnarray} 
\dot{\rho}_{11}= &&  -\dot{\rho}_{22}=  
-2\,\frac{H}{\hbar}\,\mbox{Im}\,\rho_{12}
         +\rho_{11}\rho_{22}\, \frac{2\Delta I}{S_I}\, [I(t)-I_0],  
        \label{Bayes1}\\ 
 {\dot\rho}_{12}= &&  \imat\, \frac{\varepsilon}{\hbar }\,\rho_{12}+ 
        \imat \, \frac{H}{\hbar } \, (\rho_{11}-\rho_{22})
        \nonumber \\ 
&& {}  -( \rho_{11}-  \rho_{22})  \frac{\Delta I}{S_I} \, 
[I(t)-I_0]\, \rho_{12} -\gamma \rho_{12} \, ,      
        \label{Bayes2} 
        \end{eqnarray}
where $\varepsilon$ and $H$ are, respectively, the energy asymmetry 
and tunneling strength of the TLS [the TLS Hamiltonian is 
${\cal H}_{TLS}=(\varepsilon /2) (c_1^\dagger  c_1 - c_2^\dagger c_2) 
+ H (c_1^\dagger c_2 +c_2^\dagger c_1)$], and $\gamma =\gamma_d+\gamma_e$ 
is the dephasing rate due to detector nonideality and coupling
with environment. Theoretically, $\gamma_d =0$ 
when TLS is measured by a QPC; however, if instead of QPC we use a 
single-electron transistor, then dephasing $\gamma_d$ is always 
significant.\cite{Kor-meas2} 

  Notice that the ensemble dephasing rate $\Gamma =\gamma +(\Delta I)^2/4S_I$
is larger than $\gamma$ because of different evolution of the ensemble
members due to random $I(t)$. Individual realizations can be simulated  
using the formula 
                \begin{equation}
I(t) -I_0 =  (\rho_{11} -\rho_{22}) \, \Delta I/2 +\xi (t) , 
        \label{I(t)}\end{equation}
where $\xi (t)$ is the pure white noise with spectral density $S_\xi =S_I$. 
If Eqs.\ (\ref{Bayes1})--(\ref{Bayes2}) are averaged over $\xi(t)$
(we use Stratonovich definition for stochastic differential equations),
then we get usual ensemble-averaged equations for TLS evolution 
(terms proportional to $\Delta I$ will disappear and $\gamma$ will be
replaced by $\Gamma$). 

        It is natural to characterize the effect of extra dephasing 
$\gamma_d$
by the detector ideality (efficiency) 
$\eta \equiv 1/[1+ \gamma_d 4S_I /(\Delta I)^2]$. 
One can show\cite{Kor-meas2,Averin} that $\eta = (\hbar/2\epsilon_d)^2$ 
where $\epsilon_d$ is the energy sensitivity of the detector including
the effect of backaction but neglecting the correlation between output
and backaction noises. So, an ideal case $\eta =1$ corresponds to a detector 
with quantum-limited sensitivity. 

        To realize a feedback loop (Fig.\ \ref{schematic}), we can monitor 
the TLS evolution using the detector current $I(t)$ plugged into Eqs.\
(\ref{Bayes1})--(\ref{Bayes2}). Then the TLS state is compared 
with the desired state, and the difference signal is used to
control the TLS parameters $H$ and/or $\varepsilon$. In the example
studied in this paper the feedback loop is designed to stabilize
the quantum oscillations of the state of a symmetric TLS ($\varepsilon =0$),
so the desired evolution is $\rho_{11}(t)=1-\rho_{22}(t)=
[1+\cos (\Omega t)]/2$, $\rho_{12}(t)=\rho_{21}^*(t)=
\imat \sin (\Omega t)/2$, where the frequency is 
$\Omega = (4H^2+\varepsilon^2)^{1/2}/\hbar = 2H/\hbar$. 
As a difference (``error'') signal we use the phase difference $\Delta \phi$
between the desired value $\phi_0(t)=\Omega t$ and the monitored value 
$\phi (t) \equiv \arctan \{ 2\, \mbox{Im} \rho_{12}(t)/
[\rho_{11}(t)-\rho_{22}(t)]\}$. This difference is used to control 
the TLS parameter $H$ (controlling the barrier height of DQD);   
we assume a linear dependence: $H_{fb}=H[1-F\times \Delta \phi]$,
where $F$ is the dimensionless feedback factor. 

In this paper we neglect
additional time delay\cite{Kor-meas2} in the feedback network, however, 
we take into account the finite bandwidth of a line carrying detector
current (that is the critical parameter for a possible experiment). 
More specifically, we average the current $I(t)$ with a rectangular
window of duration $\tau_a$, 
$I_a(t) \equiv \tau_a^{-1} \int_{t-\tau_a}^t I(t')dt'$, 
before plugging it into Eqs.\ (\ref{Bayes1})--(\ref{Bayes2}), so 
that the ``available'' density matrix $\rho_a(t)$ differs from the
``true'' density matrix $\rho (t)$. 
Also, to compensate for the corresponding implicit time delay, 
we use $\Delta \phi = \phi_a - \Omega (t-\kappa\tau_a)$ with
$\kappa =1/2$ (we tried various $\kappa$ and found that $\kappa =1/2$
provides the best operation of the feedback loop).

\begin{figure} [t] 
\centerline{ 
\epsfxsize=2.8in 
\epsfbox{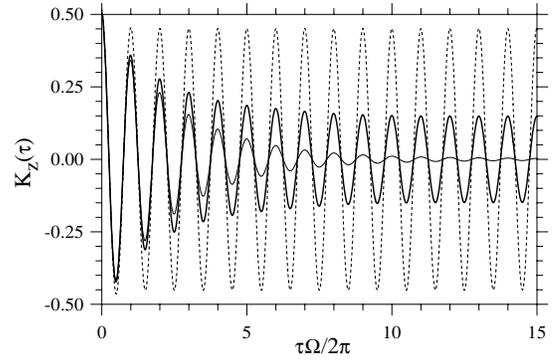}
} 
\vspace{0.4cm} 
\caption{Correlation function $K_z(\tau)$ of the TLS quantum oscillations 
for ${\cal C}=1$ and feedback factors $F=0$ (thin solid line), 
0.05 (thick solid line), and 0.5 (dashed line). Nondecaying oscillations
are due to synchronization by the feedback. 
 }
\label{K(tau)}\end{figure}

        Let us start with the case of ideal detector, $\eta =1$, 
absence of extra environment, $\gamma_e=0$, and infinite bandwidth, 
$\tau_a=0$. 
Figure \ref{K(tau)} shows numerically calculated correlation function 
$K_z (\tau) \equiv \langle z(t+\tau) z(t)\rangle$ where 
$z\equiv \rho_{11}-\rho_{22}$, for several feedback factors:
$F=0,$ 0.05, and 0.5. 
The curves are obtained using Monte Carlo simulation\cite{Kor-PRB,Kor-meas2} 
of the measurement process for moderately weak coupling between TLS 
and the detector: ${\cal C}\equiv \hbar (\Delta I)^2/S_IH=1$ (notice
that the $Q$-factor of oscillations\cite{Kor-sp} is equal to $8/{\cal C}$,
so ${\cal C}=1$ is still a weak coupling). In absence of feedback 
($F=0$) the correlation function decays to zero (Fig.\ \ref{K(tau)}) 
while for finite feedback factor the correlations remain for indefinitely 
long time (of course, assuming perfect reference oscillator which determines 
desired evolution). The nondecaying correlations show that the quantum 
feedback loop really provides the synchronization of quantum oscillations.

The degree of synchronization depends on the feedback factor $F$.
One can see that for a moderate value of $F=0.5$ the synchronization is
already very good [the ideal case would be 
$K_z(\tau )= \cos (\Omega \tau )/2$]. 
For the case of moderate or good synchronization 
(${\cal C}/16F \lesssim 1$, ${\cal C}/8 \ll 1$) 
we have derived the following analytical expression: 
        \begin{equation}
K_z(\tau )=\frac{\cos \Omega \tau}{2} \,  \exp \left[
\frac{{\cal C}}{16F} \,
\left( e^{-2FH\tau /\hbar} -1\right) \right] . 
        \label{K_z}\end{equation}
   The correlation function $K_I(\tau )\equiv \langle I(t+\tau )I(t)\rangle$
of the detector current $I(t)$ have somewhat similar dependence, however, 
it also has the decaying contribution\cite{Kor-sp} due to correlation 
$K_{z\xi}$ 
and $\delta$-function contribution due to detector noise. The analytical 
result, 
        \begin{eqnarray}
K_I(\tau )= && \frac{S_I}{2}\, \delta (\tau) +\frac{(\Delta I)^2}{4} \,
\frac{\cos (\Omega \tau )}{2} \, \left(1 +e^{-2FH\tau /\hbar} \right) 
\nonumber \\
&& \times \, \exp \left[ ({\cal C}/16F)\left( e^{-2FH\tau /\hbar}-1\right) 
\right]   ,
        \label{K_I}\end{eqnarray} 
agrees well with the Monte Carlo results. 

        The spectral density $S_{I}(\omega )$ of the detector current 
can be obtained as a Fourier transform of $K_I(\tau )$. While in absence
of feedback the quantum oscillations in TLS can provide only a moderate
peak of $S_I(\omega )$ around frequency $\Omega$ (the peak height 
cannot be larger than 
4 times the noise pedestal\cite{Kor-sp}), the feedback synchronization 
leads to the appearance of a $\delta$-function at the frequency of desired 
oscillations. (In principle the desired frequency can differ a little 
from $\Omega$; however, in this case the performance of the feedback
loop worsens.)

      Besides the correlation function and spectral density, we have  
studied one more characteristic, $D$, of the synchronization degree. 
We define $D$ as the average scalar product of the unity-length vector 
on the Bloch sphere corresponding to the desired state and the vector 
corresponding to the actual state of TLS. The equivalent definition is 
$D\equiv 2 \langle \mbox{Tr} \rho \rho_d \rangle -1$, where $\rho_d$ is the 
density matrix of the desired pure state. Perfect synchronization corresponds
to $D=1$. It is simple to show that in the case of weak coupling and 
unshifted desired frequency, $D^2/2$ is equal to the 
nondecaying amplitude of $K_z(\tau )$ dependence.

\begin{figure} [t] 
\centerline{
\epsfxsize=3.0in 
\epsfbox{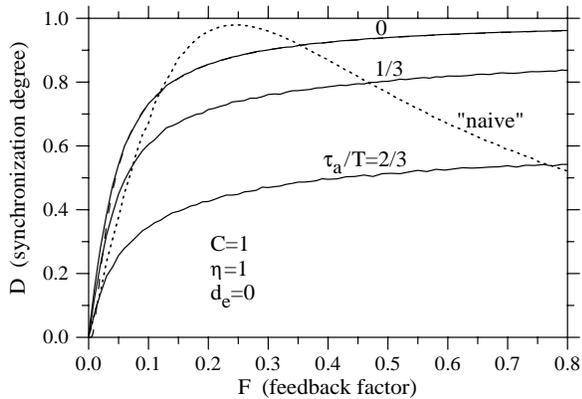}
} 
\vspace{0.4cm} 
\caption{Synchronization degree $D$ as a function of feedback factor $F$ 
for several values $\tau_a$ of detector signal averaging: $\tau_a/T=0$, 
1/3, and 2/3, where $T=2\pi/\Omega$. Dashed line $D=\exp (-{\cal C}/32F)$
almost coincides with the upper curve. 
 Dotted line corresponds to ``naive'' feedback with $\tau_a= T/10$. 
 }
\label{D(F)}\end{figure}

     Upper solid line in Fig.\ \ref{D(F)} shows the dependence 
of $D$ on the feedback factor $F$ for ${\cal C}=1$ and $\tau_a=0$. 
One can see that $D$ is proportional
to $F$ for small $F$ (``soft'' onset of synchronization) and $D$ 
is asymptotically approaching 1 at large $F$. Using Eq.\ (\ref{K_z}) 
it is simple to obtain expression $D = \exp (-{\cal C}/32F)$ which is
very close to numerical results for moderate 
and good synchronization (dashed line in Fig.\ \ref{D(F)}). 

      Finite available bandwidth of the detector current $I(t)$ 
(finite averaging time $\tau_a$ in our formalism) worsens the
performance of the quantum feedback loop. The solid lines in Fig.\ \ref{D(F)} 
show the dependence of the synchronization degree $D(F)$ for 
$\tau_a/T=0$, 1/3, and 2/3, where $T=2\pi/\Omega$ is the oscillation
period. Obviously, a significant information  loss occurs 
when $\tau_a$ becomes comparable to $T$, leading to a decrease of $D$. 
The curves $D(F)$ saturate at large $F$ allowing us to introduce 
the dependence $D_{max}(\tau )$. Calculations for parameters of
Fig.\ \ref{D(F)} show pretty good synchronization, $D_{max} =0.993$, 
for $\tau_a=T/30$, while $D_{max} =0.98$ for $\tau_a =T/10$,
$D_{max} =0.92$ for $\tau_a =T/3$, and $D_{max} =0.57$ for $\tau_a =2T/3$.

\begin{figure} [t] 
\centerline{
\epsfxsize=3.0in 
\epsfbox{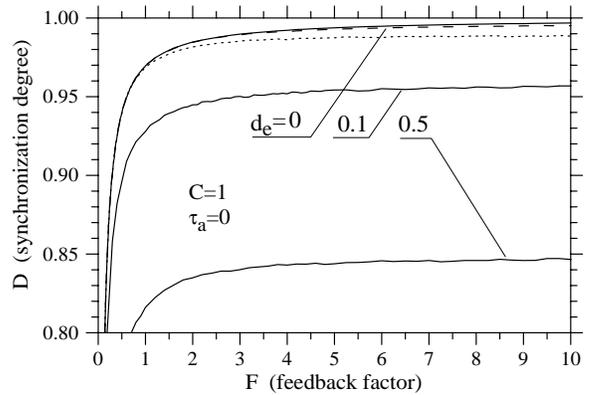}
} 
\vspace{0.4cm} 
\caption{Dependence $D(F)$ for ${\cal C}=1$, $\tau_a=0$, and several 
magnitudes of dephasing due to environment: $d_e=0$, 0.1, and 0.5. Dashed 
and dotted lines correspond to $d_e=0$ and limitation of $H_{fb}$ by 
0 and $H/2$, respectively. 
 }
\label{D(F)eta}\end{figure}

        The main potential practical importance of the quantum feedback 
is the ability to suppress the effect of TLS dephasing caused by interaction 
with environment (see Fig. \ref{schematic}). 
        Solid lines in Fig.\ \ref{D(F)eta} show the dependence 
$D(F)$ for several
magnitudes of the dephasing due to environment, $d_e=0$, 0.1, and 0.5, 
where $d_e\equiv \gamma_e /[(\Delta I)^2/4S_I]$ is the 
ratio between TLS coupling to the environment and to the detector
(we still assume an ideal detector). First of all, we see that feedback
still maintains the phase coherence of TLS for infinitely long time.
However, for finite $d_e$ the degree of synchronization $D$ saturates 
at the level less than unity. The dependence $D_{max}(d_e)$ is linear
at small $d_e$, numerically we obtained $D_{max}\simeq 1- 0.5 d_e$ 
for parameters of Fig.\ \ref{D(F)eta}. 

        Notice that the solid lines shown in Figs.\ \ref{D(F)} and 
\ref{D(F)eta} are calculated assuming the TLS feedback control 
$H_{fb}=H[1-F\times \Delta \phi]$ even when $H_{fb}$ becomes negative
(this is also an assumption for analytical results). To eliminate 
this unphysical assumption we have also performed numerical calculations 
with restriction $H_{fb}>0$ and with restriction $H_{fb}>H/2$. 
This leads to rather minor modifications of the presented curves (dashed 
and dotted lines in Fig.\ \ref{D(F)eta} show the results for $d_e=0$ and 
$\tau_a=0$). However, important difference is that $D(F)$ goes down at
large $F$, so the optimum $D_{max}$ is achieved at some finite value of $F$.
More detailed study of this problem will be presented elsewhere. 

        Besides the discussed feedback based on $\Delta \phi$ calculation,
we have also studied much simpler feedback loop in which
$H_{fb}(t)/H-1=F\{2[I_a(t)-I_0]/\Delta I -  \cos [\Omega (t-\tau_a/2)]\}
\, \sin [\Omega (t-\tau_a/2)]$ 
(in an experiment it would eliminate the necessity to solve fast the 
Bayesian equations). 
Quite surprisingly, such ``naive'' feedback can also provide a good phase 
synchronization of quantum oscillations if $F/{\cal C}$ is close to 1/4 
(see dotted line in Fig.\ \ref{D(F)}). 
 However, it requires more careful choice of $F$ and $\tau_a$ than for 
the Bayesian feedback, and also suffers more significantly from the 
restriction on $H_{fb}$ variation. 
The results in more detail will be discussed elsewhere.

Experimentally, besides the realization of quantum feedback control 
of a DQD continuously measured by a QPC, one can also think about the TLS
based on a single-Cooper-pair box measured by a single-electron transistor
(see discussion in \cite{Kor-meas2}). This realization can be preferable
because of a rapid progress of metallic single-electronics technology. 
However, the problems are high output impedance of the single-electron
transistor and its significant nonideality as a quantum detector. 
The third potential realization can be based on SQUIDs. For any realization
the major problem is bandwidth: the feedback should be faster than
the TLS dephasing due to environment. Because of that, the quantum feedback
of a TLS should probably be attempted only after the realization of recently
proposed Bell-type two-detector correlation experiment.\cite{Kor-exp}

The authors thank A. J. Leggett and G. J. Milburn for useful discussions. 
The work was supported by NSA and ARDA under ARO grant DAAD19-01-1-0491.

\end{document}